\documentclass[aps,prb,superscriptaddress,twocolumn,showpacs,amsmath,amssymb,reprint]{revtex4-2}

\usepackage{graphicx}% Include figure files
\usepackage{dcolumn}% Align table columns on decimal point
\usepackage{bm}% bold math
\usepackage{lipsum}
\usepackage{chemformula}
\usepackage[T1]{fontenc}
\usepackage[colorlinks=true, allcolors=blue]{hyperref}
\usepackage{amsfonts}
\usepackage{algorithm2e}
\usepackage[utf8]{inputenc}
\usepackage[T1]{fontenc}
\usepackage{mathptmx}
\usepackage{ulem}
\begin{document}

%\preprint{APS/123-QED}

\title{Structural phase transition in monolayer gold(I) telluride: From a room-temperature topological insulator to an auxetic semiconductor}

\author{Xin Chen}
\affiliation{Department of Physics and Astronomy, Uppsala University, Box 516,
751\,20 Uppsala, Sweden}
\author{Raquel Esteban-Puyuelo}
\affiliation{Department of Physics and Astronomy, Uppsala University, Box 516,
751\,20 Uppsala, Sweden}
\author{Linyang Li}
\email{linyang.li@hebut.edu.cn}
\affiliation{School of Science, Hebei University of Technology, Tianjin 300401, China}
\author{Biplab Sanyal}
\email{biplab.sanyal@physics.uu.se}
\affiliation{Department of Physics and Astronomy, Uppsala University, Box 516,
751\,20 Uppsala, Sweden}
\date{\today}% It is always \today, today,

\begin{abstract}
Structural phase transitions between semiconductors and topological insulators have rich applications in nanoelectronics but are rarely found in two-dimensional (2D) materials. In this work, by combining ab initio computations and evolutionary structure search, we investigate two stable 2D forms of gold(I) telluride (Au$_{2}$Te) with square symmetry, noted as s(I)- and s(II)-Au$_{2}$Te. s(II)-Au$_{2}$Te is the global minimum structure and is a room-temperature topological insulator. s(I)-Au$_{2}$Te is a direct-gap semiconductor with high carrier mobilities and unusual in-plane negative Poisson’s ratio. Both s(I) and s(II) phases have ultra-low Young’s modulus, implying high flexibility. By applying a small tensile strain, s(II)-Au$_{2}$Te can be transformed into s(I)-Au$_{2}$Te. Hence, a structural phase transition from a room-temperature topological insulator to an auxetic semiconductor is found in the 2D forms of Au$_{2}$Te, which enables potential applications in phase-change electronic devices. Moreover, we elucidate the mechanism of the phase transition with the help of phonon spectra and group theory analysis.
\end{abstract}

\maketitle

%\tableofcontents

\section{Introduction}
Graphene has been proved or predicted to have a lot of exciting properties since its first realization, such as large Young’s modulus, high mobility, and quantum spin Hall (QSH) effect \cite{Novoselov666,Novoselov2005Two-dimensionalGraphene,Zhang2005ExperimentalGraphene,PhysRevLett.95.226801}, to name a few. QSH effect is observed in two-dimensional topological insulators (2DTIs), which is a new state of matter with time-reversal symmetry protected edge states \cite{RevModPhys.83.1057,PhysRevLett.96.106802,PhysRevLett.95.226801,RevModPhys.82.3045}. Though the QSH effect was first predicted in graphene by Kane and Mele \cite{PhysRevLett.95.226801}, it has not been observed in experiments due to the ultra-weak spin-orbit coupling (SOC) effect in carbon systems. In subsequent works, the QSH effect was observed in HgTe/CdTe \cite{Konig766} and InAs/GaSb \cite{PhysRevLett.107.136603} quantum wells at ultra-low temperatures (1.4 K). Very recently, the QSH effect has been achieved in monolayer WTe$_{2}$ \cite{nphys4091}, at about 100 K, which is a significant improvement but still much lower than the room temperature. Till now, many other materials have been theoretically predicted to be room-temperature 2DTIs \cite{Qian1344,PhysRevB.102.075306,kou2015tetragonal,PhysRevB.93.165401,Chen2018Nature}. Most of the predictions are not verified by experiments yet, but we can expect them to be realized in future.

Phase transition materials (PTMs) are the materials exhibiting several stable crystalline phases with distinct properties and similar total energy.  PTMs have attracted elated research interest due to the promise for the applications in electronic and optical devices and sensors \cite{doi:10.1021/acsnano.6b00148,Qian1344,Wang2016,Wuttig2007,Wang2017}. Among the reported PTMs, the 3D metal-insulator PTMs, VO$_{2}$ and TaS$_{2}$ have been used as the channel materials for fabricating phase transition devices \cite{shukla2015steep,hollander2015electrically}. The 2D metal-insulator phase transition in MoTe$_{2}$ is also realized by thermal means \cite{Keum2015,VELLINGA1970299}, electrostatic doping\cite{doi:10.1021/acsnano.6b00148}, electrostatic gating \cite{Li2016,Wang2017,doi:10.1002/aelm.201900964}, and photoexcitation \cite{Peng2020}. Moreover, semiconductor-TI PTMs, as a new group of materials, have attracted much attention since they are an excellent platform to manipulate the topological properties of 2D materials and realize Majorana bound modes \cite{Peng2020,sajadi2018gate,fatemi2018electrically}. However, as a new class of materials, 2DTIs have rarely been reported as a stable phase of PTMs in experimental or theoretical works\cite{zhou2020normal,zhang2016semiconductor,xiao2019phase,Li2016}.

M$_{2}$X (M=metal, X=nonmetal) monolayers are a new class of 2D materials that have been paid a lot of attention\cite{doi:10.1021/ja508154e,doi:10.1021/jacs.7b06296}. The 2D forms of group \textbf{I}B-\textbf{VI}A compounds have numerous M$_{2}$X phases. 2D $\beta$-Cu$_{2}$S and $\gamma$-Cu$_{2}$S sheets have been synthesized by Romdhane et al. and Li et al.\cite{doi:10.1002/smll.201400444,doi:10.1002/adma.201602701} $\alpha$-Ag$_2$S sheets have been synthesized by Feng et al. via liquid-phase exfoliation method.\cite{doi:10.1002/adom.201901762}  Zhu et al. synthesized 2D colloidal Cu$_2$Se using the Langmuir-Blodgett (LB) method.\cite{doi:10.1021/acs.chemmater.6b01246} 2D Cu$_2$Te was synthesized by Qian et al., using molecular beam epitaxy (MBE) method.\cite{Qian_2019} Besides the experimental studies, theoretical works based on density functional theory and global structure search \cite{PhysRevB.82.094116,doi:10.1063/1.4769731} have predicted several monolayer M$_{2}$X with lower energy than the experimentally synthesized materials\cite{C8NH00216A,doi:10.1021/acs.jpclett.0c00613,doi:10.1021/acs.jpclett.9b01312,doi:10.1021/acs.nanolett.8b04761}. In a recent article, the 2D forms of M$_{2}$X (X=Cu, Ag, Au; X=S, Se) materials have been discussed \cite{doi:10.1021/acs.jpclett.0c00613}. It is found that there are two kinds of structures with square-symmetry, i.e., P4/nmm group structure s(I), and P4212 group structure s(II). For copper(I) sulfide and selenide, s(II) structures are energetically favored, while for silver(I) and gold(I) compounds, s(I) structures are favored. Due to the absence of energy barrier between s(I) and s(II) phases in most of these materials, the system will always go to one of the phases. As an exception, s(II)-Au$_{2}$Se is dynamically stable, but s(I) and s(II) phases are both direct-gap semiconductors, which make them not suitable as a semiconductor-TI PTM. Nevertheless, that inspires us to investigate other M$_{2}$X materials with stronger relativistic effects.

In this work, based on global structure search using evolutionary algorithm \cite{LYAKHOV20131172,doi:10.1063/1.2210932}, we have predicted two stable phases of Au$_{2}$Te, s(I) and s(II). The energy difference between these two phases is as low as 4 meV/atom. We have proposed structural phase transition from s(II) to s(I)-Au$_{2}$Te by external tensile strain and a reverse phase transition by chemical means at low temperature. We further investigated the influence of electric field in the phase transition, and it is found that with the application of electric field larger than 0.7 V/\AA, s(I) has lower energy than s(II), which makes the s(II)$\longrightarrow$s(I) phase transition possible with the assistance of heat. Moreover, it is found that they are quite different in their electronic and mechanical properties. s(I)-Au$_{2}$Te is a direct-gap semiconductor with high electron and hole mobilities of $3.45 \times 10^{4} cm^{2}/(Vs)$ and $6.47 \times 10^{3} cm^{2}/(Vs)$, while s(II) phase is a TI with a large nontrivial band-gap of 28.4 meV. Both of the two structures have ultra-low Young’s modulus, showing extremely high flexibility. But by applying strain, only s(I) phase shows unusual negative Poisson’s ratio (NPR). Finally, we computed the Raman spectra of them and have shown that the peak at about 65.9 cm$^{-1}$ is only observed in s(II) phase, which can be an excellent method to distinguish the two similar phases.

\section{Computational details}
We performed the global structure search with the evolutionary algorithm based code USPEX \cite{doi:10.1063/1.2210932,GLASS2006713,PhysRevB.87.195317,doi:10.1021/ar1001318,LYAKHOV20131172} to obtain low energy stable structures. In the structure search, the unit cell is chosen to contain four Au atoms and two Te atoms, and the initial thickness of the region is set as 0-4 \AA. The population size is set to 30. The structure search is converged if the ground state structure did not change for 10 generations. All the first-principles calculations are performed by using the density functional code VASP\cite{002230939500355X,PhysRevB.54.11169}. The generalized gradient approximation in the form of  Perdew, Burke, and Ernzerhof (PBE) \cite{PhysRevLett.77.3865} is used for the exchange-correlation potential. For the calculations of 2D materials, the out-of-plane interaction is avoided by taking a vacuum of more than 20 \AA. The energy cutoff of the plane waves is set to more than 450 eV. The tolerance for energy convergence is set to be less than 10$^{-5}$ eV. We optimized the structures until the force on each atom becomes smaller than 0.001 eV/\AA. The Brillouin zone (BZ) is sampled by using the Monkhorst-Pack grid denser than $2 \pi \times 0.033$ \AA$^{-1}$. For examining the dynamical stability of structures, phonon spectra are computed using the PHONOPY code \cite{phonopy}. In the calculations of the electronic band structures, Heyd-Scuseria-Ernzerhof (HSE) hybrid functional within the framework of HSE06 \cite{Krukau2006} is employed. Post-Processing of some calculations are performed by using VASPKIT \cite{wang2019vaspkit} and VESTA \cite{Mommako5060}.

\section{Results and discussion}
\subsection{Structure and stability}
Among the hundreds of structures generated in the evolutionary structural search for Au$_{2}$Te, the square symmetry structures s(I) and s(II), as shown in Fig. \ref{figure1} (a), are found to be energetically favored, as shown in Fig. S1 in the Supplemental Material (SM) \cite{SuppMater}. S(I)- and s(II)-Au$_{2}$Te is the second-lowest and the lowest energy structures, showing P4/nmm and P4212 space group symmetry, respectively. In detail, there are four Au atoms sandwiched by two Te atoms in one s(I/II)-Au$_{2}$Te unit cell. The relationship between s(I) and s(II) phases is like that between 1T and 1T$^{\prime}$ phases of MoTe$_{2}$, i.e., the symmetry of the structures is reduced by distortion. The critical parameters of the two structures are listed in Table \ref{table1}. The lattice constant of s(I) phase is 4.3$\%$ larger than that of s(II), while the thickness of the former is 3.1$\%$ smaller than the latter. The electron localization functions (ELFs) are shown in Fig. \ref{figure1} (b), where it is observed that the electrons are mainly localized near Te atoms, but the electron density near Au is much more than that near the Cu and Ag atoms in s(II)-Cu$_{2}$S and s(I)-Ag$_{2}$S monolayers. This can be explained by the much lower electronegativity of tellurium atoms compared to that of sulfur atoms.

\begin{figure*}[hbtp]
\includegraphics[scale=0.8]{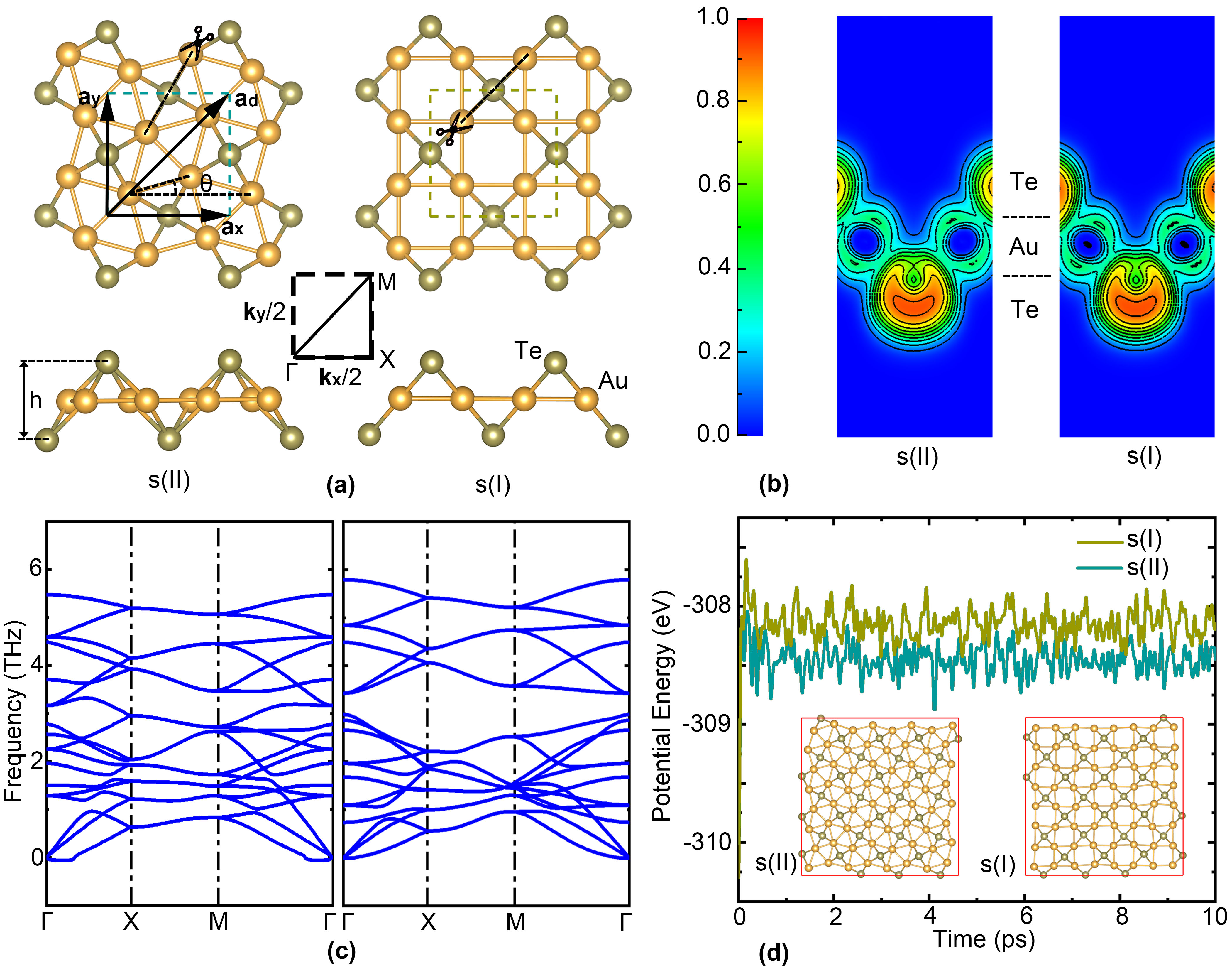} 
\caption{(a) The top and side views of the atomic structures of s(II)- and s(I)-Au$_{2}$Te monolayers with their primitive cell marked by dashed square. $\mathbf{a_{x}}$, $\mathbf{a_{y}}$, $\mathbf{a_{d}}$, h and $\theta$ are the lattice vectors in the x-, y-, and the diagonal-directions, the heights, and the distortion angles of s(I/II)-Au$_{2}$Te. The middle dashed black square is the first BZ of s(I/II)-Au$_{2}$Te monolayers, with the high symmetric points indicated. (b) The contour maps of ELFs of s(II)- and s(I)-Au$_{2}$Te, which are sliced on Au-Te-Au planes marked with scissors in (a).  (c) The phonon dispersion of s(I/II)-Au$_{2}$Te. (d) The potential energy change as a function of simulation time at 300 K in BOMD simulations along with the final structures.
}
\label{figure1}
\end{figure*}

To study the energetic stability and compare to other 2D materials, we compute the cohesive energy following the expression
\begin{equation}
\Delta E=\frac{ E_{Au_{2}Te}-(n_{Au} \times E_{Au}+n_{Te} \times E_{Te}) }{n_{Au}+n_{Te}}.
\label{eq1}
\end{equation}
In this formula, E$_{Au_{2}Te}$ is the total energy of s(I/II)-Au$_{2}$Te, E$_{Au}$, and E$_{Te}$ are the energies of Au and Te atoms. The values of $\Delta E$ are listed in Table \ref{table1}. For comparison, the values of the synthesized germanene and stanene computed, as -3.26 eV/atom and -2.74 eV/atom, respectively \cite{doi:10.1021/ja513209c}. The low cohesive energy implies that the synthesis of s(I/II)-Au$_{2}$Te is highly probable. 

To further strengthen aspect of the energetic stability of s(I/II)-Au$_{2}$Te, we performed a  variable-composition structure search with a unit cell containing up to 6 Au or Te atoms. The objective function was defined as $E_{f}=(E(Au_{x}Te_{y})-x E(Au)-y E(Te))/(x+y)$, in which  $E(Au)$ and $E(Te)$ are the total energies of the ground state structures of 2D Au and Te structures, respectively. From the convex hull diagram, as shown in Fig. S5 of the SM, it is confirmed that s(II)-Au$_{2}$Te is one of the most stable structures in the 2D Au-Te system.

%=======================================================
\begin{table}[htbp]
\begin{center}
\caption{The lattice parameters (a$_{x}$ and a$_{d}$, in \AA), slab thicknesses (h, in \AA), distortion angles ($\theta$, in~$^{\circ})$, cohesive energies ($\Delta E$, in eV/atom), and stiffness tensors (C11, C12, and C66, in N/m) of s(I/II)-Au$_{2}$Te monolayers.}
\label{table1}
\tabcolsep2pt
\arrayrulewidth0.5pt
\begin{tabular}{|c|c|c|c|c|c|c|c|c|}
\hline
\hline Phase&a$_{x}$&a$_{d}$&h&$\theta$&$\Delta E$&C11&C12&C66\\
\hline s(I)&5.85&8.27&3.44&0&-2.893&23.460&5.953&12.301\\
\hline s(II)&5.61&7.93&3.55&15.28&-2.897&34.798&19.006&14.199\\
\hline
\hline
\end{tabular}
\end{center}
\end{table}
%=======================================================

The dynamic stability of s(I/II)-Au$_{2}$Te is confirmed by the density functional perturbation theory (DFPT) calculations and ab initio Born-Oppenheimer molecular dynamics (BOMD) simulations. In the DFPT calculations, we employed a $3 \times 3 \times 1$ supercell and a $3 \times 3 \times 1$ $\Gamma$-centered k-mesh. As shown in Fig. \ref{figure1}(c), in the phonon spectra of both s(I) and s(II) phases, there are no or only tiny imaginary frequencies (less than 0.1 Thz) near $\Gamma$ point. In the BOMD simulations, we have checked the structural and energetic change of the $4 \times 4 \times 1$ supercells of s(I)- and s(II)-Au$_{2}$Te at 300 K until 10 ps, with a time step of 1 fs. As shown in Fig. \ref{figure1}(d), the energies of the two phases are kept stable, and there is no obvious structural change in the final geometrical framework.

A mechanically stable 2D material should satisfy the Born-Huang criteria: the elastic modulus tensor components C$_{11}$, $C_{22}$, and $C_{66}$ should be positive, and $|C_{11}+C_{22}| > |2C_{12}$|. We fitted the curves of the energy changes U versus strains $\tau_{x y}$ using formula \cite{PhysRevB.85.125428,Zhang201416591}:
\begin{equation}
U=\frac{1}{2} C_{11} \tau_{x}^{2}+\frac{1}{2} C_{22} \tau_{y}^{2}+C_{12} \tau_{x} \tau_{y}+2 C_{66} \tau_{x y}^{2}.
\label{eq2}
\end{equation}
Considering the square symmetry, $C_{22}$ is equal to $C_{11}$, and the rest of the components of stiffness tensors are listed in Table \ref{table1}. The results meet the criteria, confirming that both phases of Au$_{2}$Te are mechanically stable. In the light of other 2D compounds of transition metal and group \textbf{IV}A elements, we suggest that the chemical vapor deposition (CVD) method could be a possible way to synthesize s(I/II)-Au$_{2}$Te  \cite{doi:10.1002/smll.201400444,doi:10.1021/nn501175k, Naylor_2017}.

%============================================
\begin{figure}[htbp]
\includegraphics[scale=0.9]{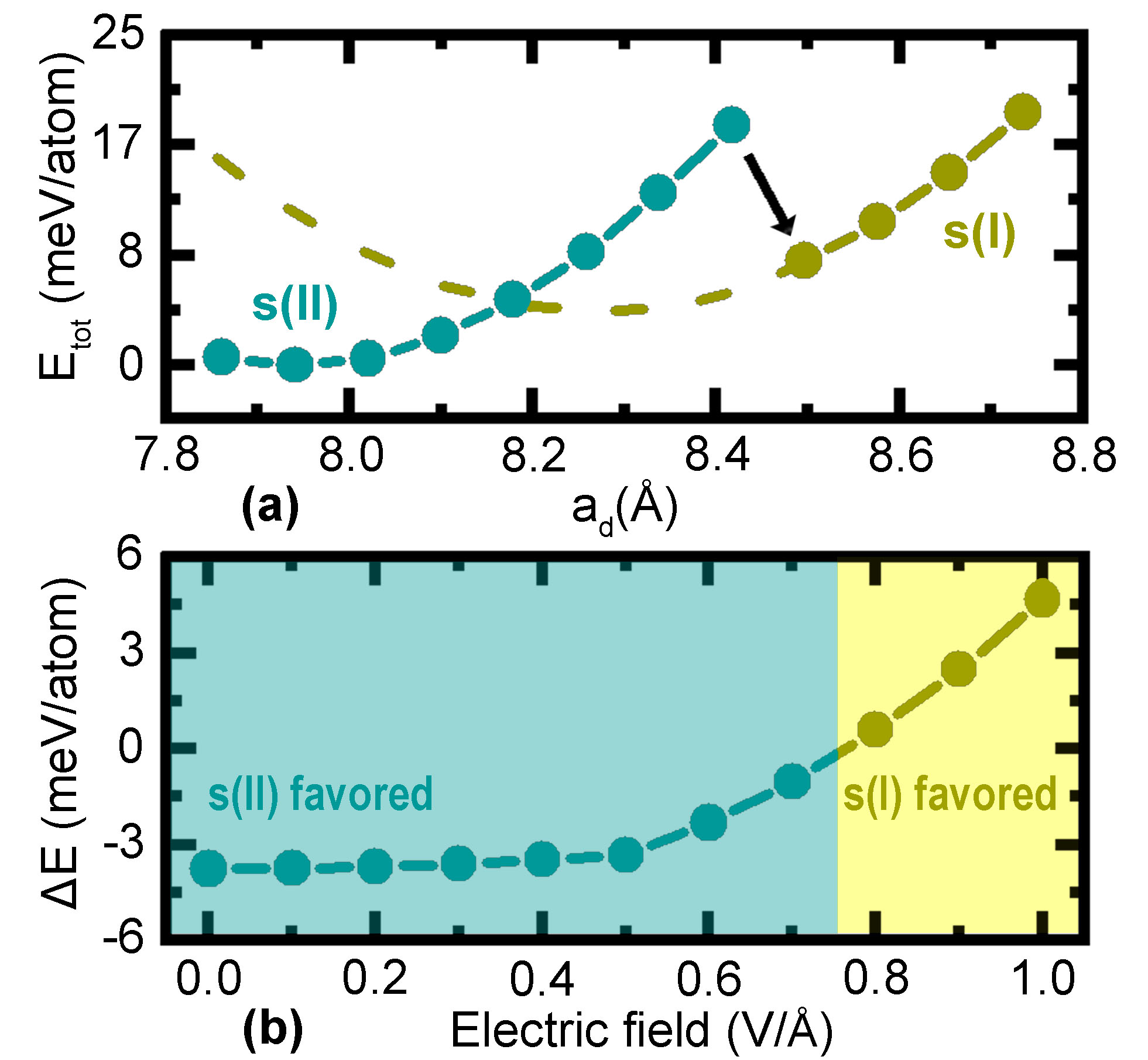}
\caption{(a) The energy change of the structure as a function of diagonal lattice constant $a_{d}$. The two local minima represent the fully relaxed s(I)- and s(II)-Au$_{2}$Te structures. (b) The energy difference between the two phases as a function of the strength of the electric field in the z-direction.}
\label{figure2}
\end{figure}
%============================================

\subsection{Structrual phase transition}
Notably, the energy difference between the two phases is just 4 meV. With similar structures and energies, the phase transition between the two phases is quite promising. Considering the obvious difference between the lattice parameters of the two phases, we investigated the feasibility of introducing the phase transition from s(II) to s(I) structure by changing the lattice parameters (applying external tensile strain), and the energy change in the process is shown in Fig. \ref{figure2} (a). Interestingly, it is found that by applying a tensile strain of above about 6.5 $\%$ along the (110)-direction, s(II) structure can be transformed into s(I) structure, with an energy barrier of about 19 meV/atom. 

\begin{figure}
\includegraphics[scale=1.0]{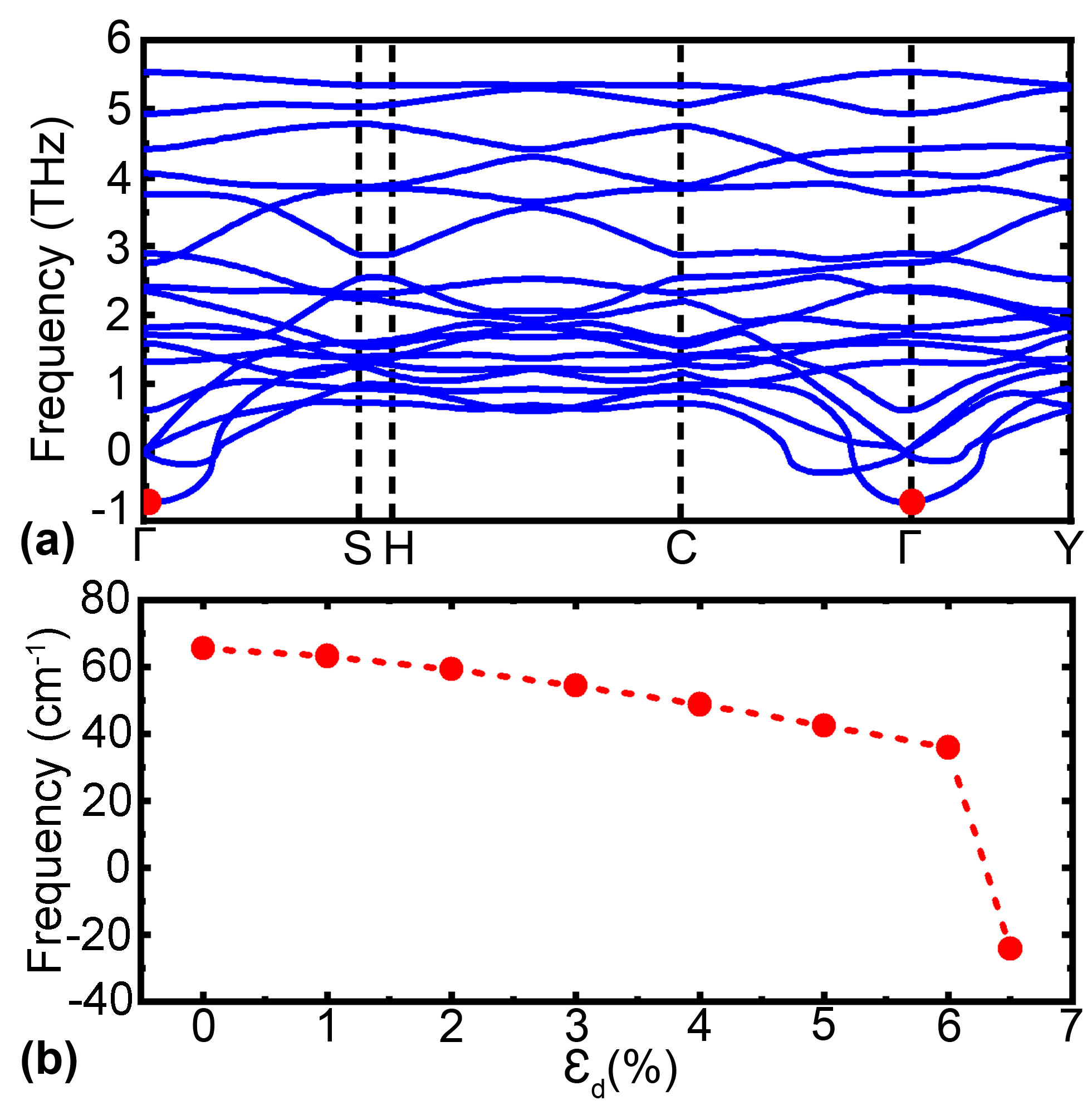}
\caption{(a) The phonon spectrum of s(II)-Au$_{2}$Te with diagonal strain $\epsilon_{d}$ of 6.5 $\%$ along the high-symmetry path as shown in Fig. S6 (c). The red dots mark the phonon softening, and the corresponding phonon eigenvector is shown in the movie file epsilon6.5.mp4 in SM. (b) The frequency of the phonon mode of the phase-transition eigenvector as a function of diagonal strain.}
\label{figure3}
\end{figure}

The s(I) phase is a more symmetric phase, belonging to the D$_{4h}$ point group. In comparison, the s(II) phase can be classified as the D$_{4}$ point group, which is a subgroup of D$_{4h}$. The difference between the two point groups is that the D$_{4}$ is not invariant under inversion symmetry, mirror plane symmetries, and therefore improper symmetries (which are a combination of a mirror plane symmetry and a rotation), whereas D$_{4h}$ has all these symmetries. To connect the symmetry arguments with eigenmodes of phonons, we further computed the phonon spectra of s(II)-Au$_{2}$Te with different diagonal strains ($\epsilon_{d}$) to investigate the phonon softening in the strain-induced phase transition process, as shown in Figs. \ref{figure3} and Fig. S6 in the SM \cite{SuppMater}. The diagonal strain lowers the symmetry from D$_{4}$ to D$_{2}$, where even fewer rotations leave the structure invariant. It is observed that with a strain of 6 $\%$, s(II)-Au$_{2}$Te is still stable. With a strain of 6.5 $\%$, there is a softened phonon mode near the $\Gamma$ point. The eigenvector of the mode at the $\Gamma$ point is shown in the attached movie file epsilon6.5.mp4 in the SM \cite{SuppMater}, which is actually the phase transition displacement from s(II) to s(I). By tracking the frequencies of this phonon mode with the special eigenvector under varying diagonal strains, as shown in Fig. \ref{figure3}(b), we have observed that the strain lowers the energy of this branch continuously until it reaches 6.5 $\%$, where the abrupt transition happens. Therefore, even if the phonon mode allows the transition to be present at all stages, it is not until it is lowered enough in energy that the system undergoes the distortion that promotes the change of phase. With a strain of 7 $\%$, the structure is transferred into the s(I) phase, and no imaginary frequencies are observed.

%=======================================================

%=======================================================

\begin{figure}[htbp]
\includegraphics[scale=0.9]{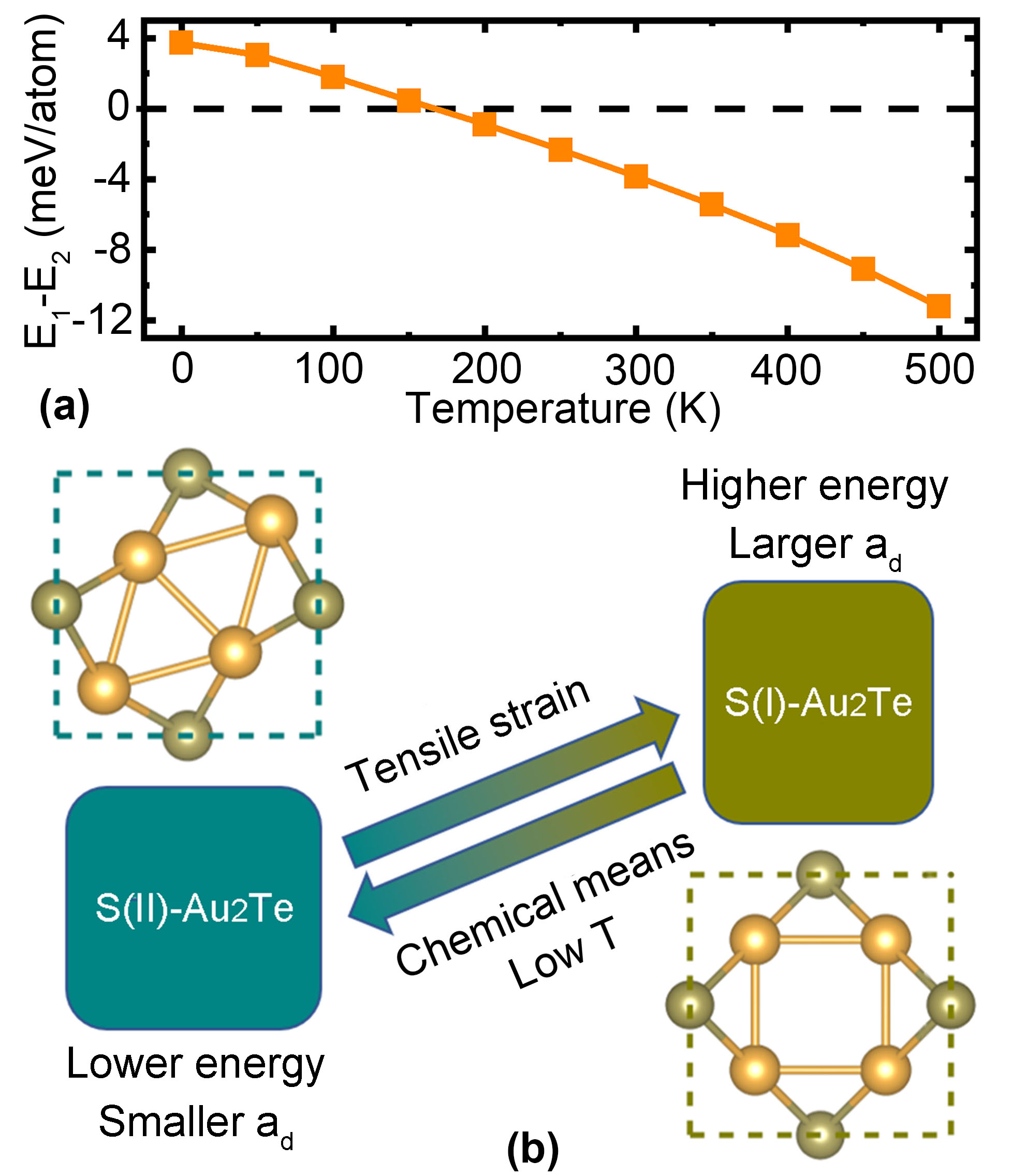}
\caption{(a) The free energy difference between s(I) and s(II) as a function of temperature. (b) The mechanism of the reversible structural phase transition is indicated by mechanical and thermal means.}
\label{figure4}
\end{figure}

Compared to tensile strain, applying external compressive strain is much harder\cite{srep16108,Zhao2015}. Therefore, to achieve the phase transition from s(I) to s(II) structure, heating the system seems to be more achievable than applying compressive strain. By using the variable cell nudged elastic band method (VC-NEB)\cite{QIAN20132111}, we investigated the phase transition mechanism between s(I) and s(II). As shown in Fig. S2 in the Supplemental Material \cite{SuppMater}, the energy barrier of the transition from s(I) to s(II) is about 7.8 meV/atom. Since the elastic constants of s(I) are softer than that of s(II), the free energy of s(I) decreases faster than that of s(II). Based on quasi harmonic approximation, we investigated the influence of temperature on the Helmholtz free energies, as shown in Fig. \ref{figure4} (a). It is found that the free energy of the s(I)-phase decreases faster than the s(II)-phase, which makes the s(I)-phase more stable (about 4 meV/atom lower) than s(II)-phase at 300 K. Thus we cannot state that a s(I)-s(II) phase transition can be achieved by thermal excitation only. Instead, decreasing the temperature and with the help of some chemical methods such as catalysis, we may achieve the phase transition from s(II) to s(I), as shown in Fig. \ref{figure4} (b).

Notably, the higher total energy of the s(II)-phase does not mean that the s(II) phase will transfer to s(I)-phase at room temperature since the energy barrier is much larger than the total energy difference (4 meV/atom) at 300K. We can see the evidence in the BOMD simulation, even when the lattice parameters are fixed. The lattice parameter of s(I)-phase is 4.3 $\%$ larger than s(I)-phase, thus if s(II)-phase was not thermally stable, the 2D structure would buckle and wrinkle, which is not observed in the final structure. Moreover, we have performed an isobaric-isothermic BOMD simulation with variable lattice parameters \cite{PhysRevLett.45.1196,doi:10.1063/1.328693}. It is found that during the simulation time up to 10 ps, the s(II) phase is always stable, and the phase transition from s(II) to s(I) is not observed. The relevant details are shown in Fig. S7 in the SM \cite{SuppMater}.

%============================================
%============================================

Moreover, the phase transition driven by the electric field has been paid extensive attention due to the excellent connection with the electronic industry \cite{Li2016,Wang2017,doi:10.1002/aelm.201900964}. In this work, by applying the external electric field in the z-direction, we studied the energetic properties of s(I) and s(II)-Au$_{2}$Te. The energy change and the energy difference between s(I) and s(II) phases are shown in Fig. \ref{figure2}(b). It is found that the energies of the two phases have similar Energy-E-field curves, but the energy of s(I) phase decreases sharper than that of s(II) phase, which makes s(I) more energetically favored. Thus with the help of the external electric field, the s(II) $\longrightarrow$s(I) phase transition can be easier. However, achieving an ultra-strong electric field is very hard in the experiment, which means that in the 2D Au$_{2}$Te case, the electric field is just an assistant method to make the phase transition easier, and it cannot drive the phase transition alone.

%=======================================================
\begin{figure*}
\includegraphics[scale=0.9]{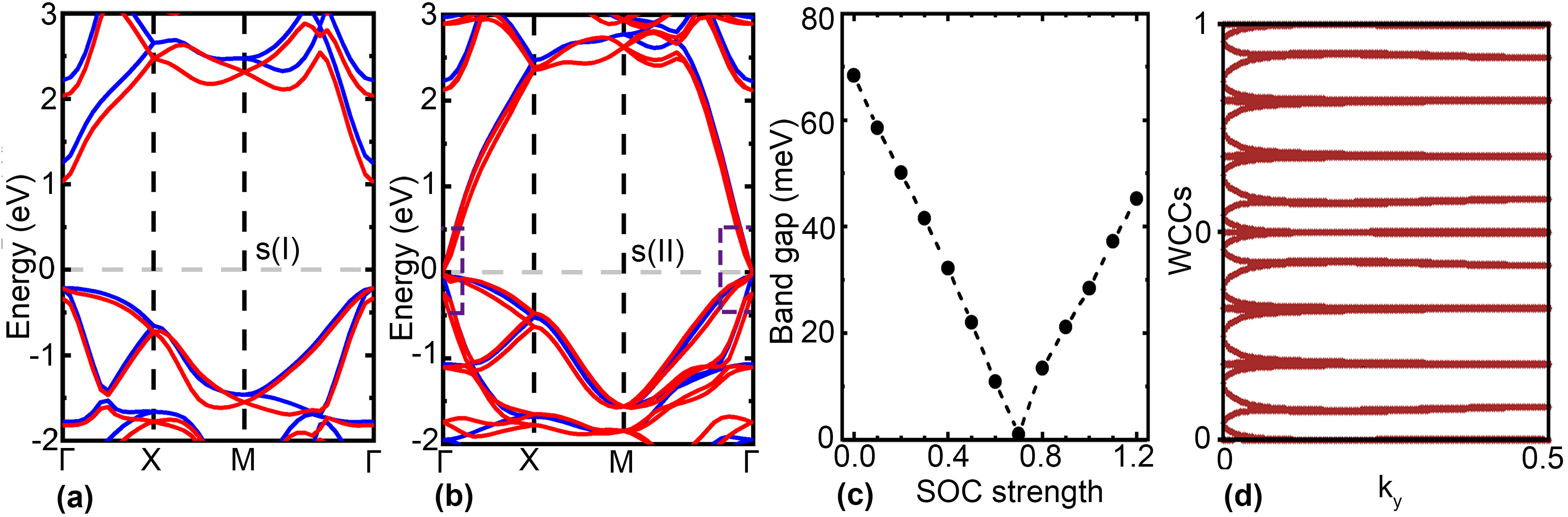}
\caption{The electronic band structures of (a) s(I)-Au$_{2}$Te and (b) s(II)-Au$_{2}$Te. The purple rectangle parts of s(II)-Au$_{2}$Te band structure are enlarged and shown in Fig. S4 in the SM \cite{SuppMater}. The contribution of each atomic orbital is shown by the local density of states (LDOS) in Fig. S2 in the SM \cite{SuppMater}. The blue and red lines are the results of HSE calculations without and with SOC, respectively. (c) shows the bang-gap of s(II)-Au$_{2}$Te as a function of SOC strength. (d) shows the evolution of Wannier charge centers.}
\label{figure5}
\end{figure*}

%=======================================================

\subsection{Electronic and topological properties}
The electronic band structures of 2D s(I/II)-Au$_{2}$Te are computed using the HSE06 method with and without the SOC effect taken into account, as shown in Fig. \ref{figure5}. s(I)-Au$_{2}$Te is a direct-gap semiconductor, with the valence band maximum (VBM) and conduction band minimum (CBM) are both at Gamma point, similar to other reported s-A$_{2}$B structures \cite{doi:10.1002/smll.201400444,doi:10.1002/adma.201602701,C8NH00216A,doi:10.1021/acs.jpclett.9b01312}. 
The relativistic effect is prominently seen in the band structures due to the large atomic mass. Without SOC, the bandgap is as large as 1.470 eV; with SOC, the bandgap is decreased to 1.252 eV.

The electronic conductivity is a crucial property for the application of semiconductors. We computed the carrier mobilities by using the deformation potential method: \cite{PhysRev.80.72,doi:10.1063/1.3665183}
\begin{equation}
\mu=\frac{e \hbar^{3} C_{11}}{k_{\mathrm{B}} \operatorname{T}m^{*} m^{*}_{\mathrm{d}}\left(E^{n}_{1}\right)^{2}},
\end{equation}
where $C_{11}$ is the 2D elastic modulus given in Eq. \ref{eq1}. $e$, $\hbar$, $k_{B}$, and $T$ are the electron charge, reduced Plank constant, Boltzmann constant, and temperature (300 K) respectively. $m^{*}$ and $m^{*}_{d}$ are the effective mass the average effective mass derived from $m^{*}_{\mathrm{d}}=\sqrt{m^{*} m^{*}_{\perp}}$, where $m^{*}_{\perp}$ is the effective mass in the perpendicular direction of the transport direction. $E^{n}_{1}$ is the deformation potential constant of band $n$, and is defined by $E^{n}_{1}=\Delta V^{n}/\left(\Delta l / l_{0}\right)$, where $\Delta V^{n}$ is the change of the edge of band $n$ under deformation $\Delta l$. The transport direction is assumed as the x-direction. The electronic transport properties are determined by the computed quantities as m$^{*}_{d,e}=m^{*}_{e}=0.082~m_{0}$, $E_{1,e}=1.433~eV$, and thus $\mu_{e}=3.45 \times 10^{4}~cm^{2}/(Vs)$ whereas for the hole transport, the corresponding quantities are computed as m$^{*}_{d,h}=m^{*}_{h}=-0.146~m_{0}$, $E_{1,h}=-2.029 ~eV$, and thus $\mu_{h}=6.466 \times 10^{3}~cm^{2}/(Vs)$. The mobilities are much larger than that of 1-H MoS$_{2}$ and black phosphorus \cite{Qiao2014,PhysRev.163.743,Radisavljevic2011}, promising potential applications in nanodevices and new generation solar cells.

On the other hand, the electronic band structures of s(II)-Au$_{2}$Te calculated without and with SOC are shown in Fig. \ref{figure5}(b). Without SOC, both VBM and CBM are at the Gamma point, with a direct bandgap of 68.4 meV. With SOC, the spin degeneracy is lifted due to the asymmetric geometry, similar to the case in monolayer III-Bi \cite{doi:10.1021/nl500206u}. The lifting is contributed by the spin-orbit interaction and the bulk inversion asymmetric structure induced Rashba and Dresselhaus spin-orbit terms in the effective Hamiltonian. The conduction band near the CBM splits into two bands, but the CBM is still at the $\Gamma$ point. In contrast, the highest two valence bands near the $\Gamma$ point split into four bands, and the VBM is moved away from the $\Gamma$ point. Between the CBM and VBM, there is an indirect bandgap of 28.4 meV.  To check if a band inversion happens between the CBM and VBM due to the SOC effect, we perform a series of band structure calculations, in which different SOC strengths ($\lambda_{SOC}$) are employed \cite{kou2015tetragonal,huang2016interface}. 

As shown in Fig. \ref{figure5} (c), with increasing SOC strengths from 0 to 0.7, the bandgap of s(II)-Au$_{2}$Te decreases. When $\lambda_{SOC}$ reaches 0.7, a gapless electronic band structure is observed. When we continue to increase the strength of SOC, the bandgap increases. The observation was also found in many other 2D materials, such as DHF GaBi-Br$_{2}$ (X=I, Br, Cl) \cite{doi:10.1007/s12274-017-1464-z},  GaBi monolayer \cite{doi:10.1021/nl500206u}, and tetragonal Bi bilayer \cite{kou2015tetragonal}, in which the band inversion changes the $\mathbb{Z}_{2}$ and induces the topologically nontrivial nature of these materials. 

To confirm the topologically nontrivial nature of s(II)-Au$_{2}$Te, we further calculated the ${\mathbb{Z}}_{2}$ topological invariant. Due to the lack of inversion symmetry in s(II)-Au$_{2}$Te, the parity criterion proposed by Fu and Kane \cite{PhysRevB.76.045302} is not sufficient to get the ${\mathbb{Z}}_{2}$ index. A further study on the topological property was carried out within the WannierTools package \cite{MOSTOFI20142309,WU2018405,PhysRevB.83.235401,PhysRevB.84.075119}. We calculated the evolution of Wannier charge centers (WCCs) as shown in Fig. \ref{figure5}(d), where ${\mathbb{Z}}_{2}=1$ can be obtained \cite{PhysRevB.83.035108}. The band inversion at the $\Gamma$ point and ${\mathbb{Z}}_{2}=1$ confirm that s(II)-Au$_{2}$Te is a nontrivial QSH insulator.
\begin{figure*}
\includegraphics[scale=0.9]{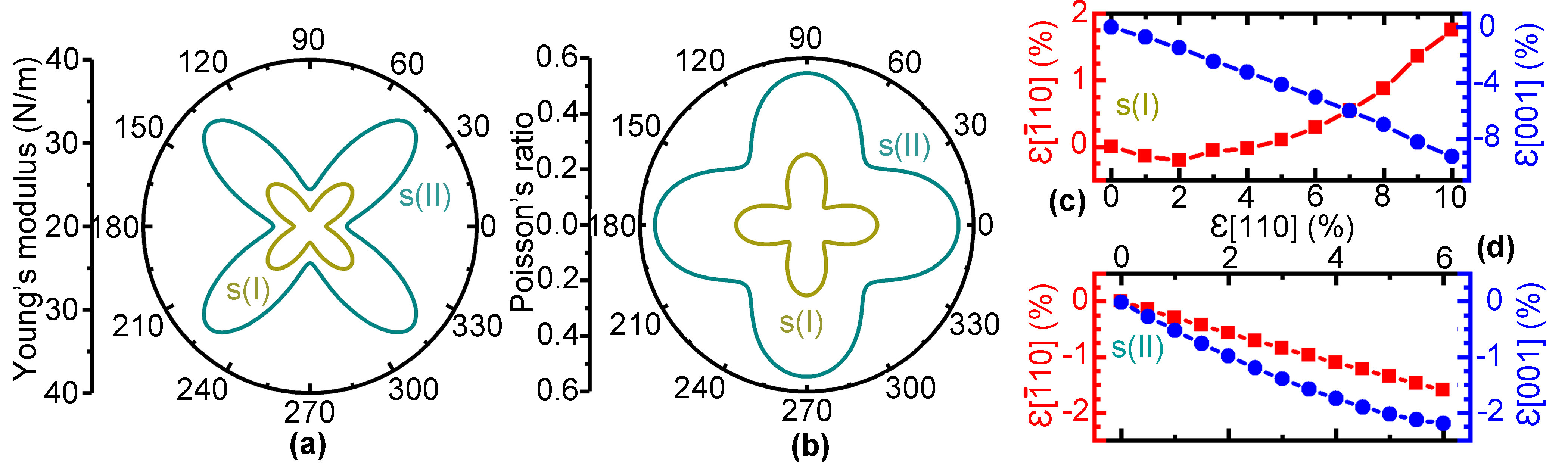}
\caption{(a) and (b) show the polar diagrams of in-plane Young's moduli and Poisson's ratios of s(I/II)-Au$_{2}$Te monolayers. $\theta=0$ and $\theta=90$ correspond to x- and y-direction, respectively. (c) and (d) show the mechanical response along [$\overline{1}$10] direction and [001] direction under strain along [110]-direction of 2D s(I)- and s(II)-Au$_{2}$Te, respectively.}
\label{figure6}
\end{figure*}

\subsection{Mechanical properties and strain-induced NPR}
The direction-dependent mechanical properties of s(I/II)-Au$_{2}$Te can be computed from the elastic parameters we got via Eq. \ref{eq2}. Using the two formulae given by Cadelano et al. \cite{PhysRevB.85.245434,PhysRevB.82.235414}, we can compute the in-plane Young’s moduli and Poisson’s ratios as:
\begin{widetext}
\begin{equation}
E_{2D}(\alpha)=\frac{C_{11} C_{22}-C_{12}^{2}}{C_{11} \sin ^{4} \alpha+C_{22} \cos ^{4} \alpha+\left(\frac{C_{11} C_{22}-C_{12}^{2}}{C_{66}}-2 C_{12}\right) \cos ^{2} \alpha \sin ^{2} \alpha}, 
\end{equation}
and
\begin{equation}
\vartheta(\alpha)=-\frac{\left(C_{11}+C_{22}-\frac{C_{11} C_{22}-C_{12}^{2}}{C_{66}}\right) \cos ^{2} \alpha \sin ^{2} \alpha-C_{12}\left(\cos ^{4} \alpha+\sin ^{4} \alpha\right)}{C_{11} \sin ^{4} \alpha+C_{22} \cos ^{4} \alpha+\left(\frac{C_{11} C_{22}-C_{12}^{2}}{C_{66}}-2 C_{12}\right) \cos ^{2} \alpha \sin ^{2} \alpha},
\end{equation}
\end{widetext} 
respectively. In which, $\alpha$ is the angle of the direction, where we define the x-direction as 0$^{\circ}$ and the y-direction as 90$^{\circ}$. As shown in Fig. \ref{figure6} (a) and (b), s(I/II)-Au$_{2}$Te have anisotropic in-plane Young’s moduli and Poisson’s ratios. The softest directions of the two phases are both along the [100]- and [010]-directions, with their Young’s moduli of 21.95 N/m (s(I)) and 24.42 N/m (s(II)). The hardest directions of them are both along the [110]- and [$\overline{1}$10]-directions, with their Young’s moduli of 26.79 N/m (s(I)) and 37.17 N/m (s(II)).  In comparison, Young’s modulus of graphene is as large as 335 N/m \cite{Lee385}. The small in-plane Young’s moduli show extraordinary flexibilities and are also observed in s-A$_{2}$B (A=Cu, Ag, Au; B=S, Se) and $\alpha$-Ag$_{2}$S monolayers \cite{doi:10.1021/acs.jpclett.0c00613,doi:10.1021/acs.nanolett.8b04761}. The minima of the direction-dependent Poisson’s ratios are both along the [110]- and [$\overline{1}$10]-directions whereas the maxima are along the [100] and [010]-directions. In each direction, the Poisson’s ratios of s(I)-Au$_{2}$Te are much smaller than that of s(II)-Au$_{2}$Te.

\begin{figure}[htbp]
\includegraphics[scale=0.9]{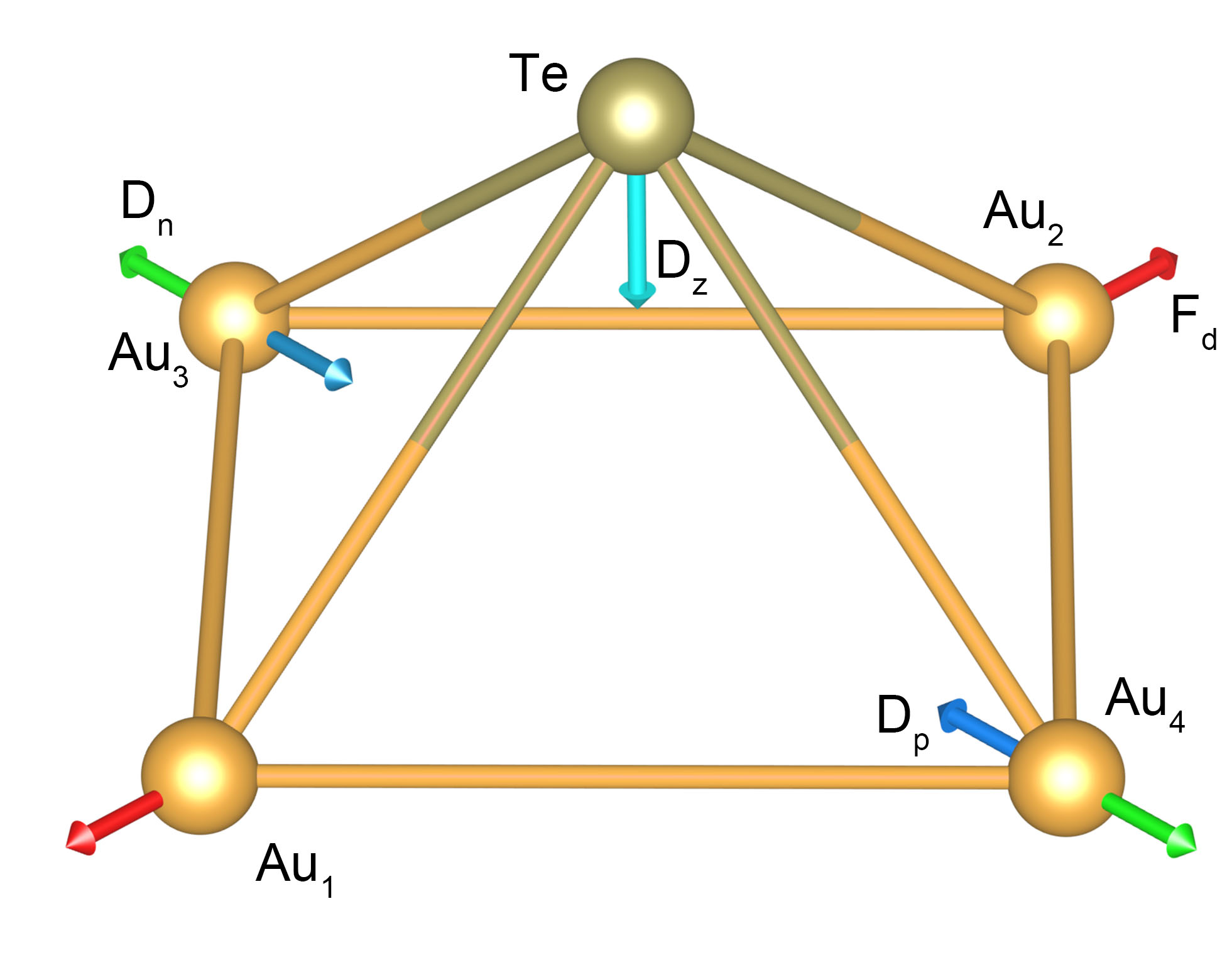}
\caption{The geometric changes under external tensile strain F$_{d}$ as indicated by the red vectors on atoms Au$_{1}$ and Au$_{2}$. The deformation of Te atom, D$_{z}$, is indicated by the light blue vector. The deformation of atoms Au$_{3}$ and Au$_{4}$ in NPR and PPR cases, D$_{n}$ and D$_{p}$, are marked by the green and the blue vectors, respectively.}
\label{figure7}
\end{figure}

Auxetic materials are characterized by its unusual NPR: when a tensile$\/$compressive strain is applied in one direction, they will expand$\/$shrink in the vertical direction. A lot of attractive properties have been found in auxetic materials, such as robust shear resistance, enhanced sound absorption, which make them quite promising in the areas of medicine, clothing materials, and tough composites \cite{PhysRevLett.84.5548,Park2013,Evans2000}. However, NPR in two-dimensional materials is rarely found \cite{Zhang201416591, doi:10.1021/acs.nanolett.6b03921, C7NR06932D, doi:10.1021/acs.nanolett.6b04180, Jiang2014, doi:10.1021/acs.nanolett.8b04761,doi:10.1021/acs.jpclett.9b00762}. In view of the fact that most of the other s-A$_{2}$B structures have NPR \cite{ doi:10.1021/acs.jpclett.0c00613}, we have also computed the in-plane and out-of-plane mechanical response of s(I/II)-Au$_{2}$Te.  

In these calculations, the strain $\epsilon[110]$ is defined as $\varepsilon[110]=a_{d}^{\prime}/a_{d}-1$, and $ a_{d}^{\prime}$ is the length of the [110]-direction diagonal of unit cell. The mechanical responses in [1$\bar{1}$0]- and [001]-directions are given by the change of the length of the [1$\bar{1}$0]-direction diagonal and the height of the unit cells, respectively. As given previously in Fig. \ref{figure2} (a), s(II) phase is kept stable when $a_{d}<8.4$, thus the mechanical response of s(II)-Au$_{2}$Te is computed in the range of $0 \le \varepsilon[110] \le 6\%$.  As shown in Fig. \ref{figure6} (d), with increasing strain $\epsilon[110]$, the mechanical responses in [1$\bar{1}$0]- and [001]-directions, $\varepsilon[1\bar{1}0]$ and $\varepsilon[001]$ are always decreasing. According to the definition, the Poisson’s ratio $\nu=-d\varepsilon_{//}/d\varepsilon_{\perp}$, in which $\varepsilon_{//}$ and $\varepsilon_{\perp}$ are the relative strain along an arbitrary direction and its perpendicular direction. Thus both the in-plane and out-of-plane Poisson’s ratios of s(II)-Au$_{2}$Te are positive. On the other hand, as shown in Fig. \ref{figure6} (c), the out-of-plane Poisson’s ratios of s(I)-Au$_{2}$Te are also always positive like s(II) phase, but the in-plane Poisson’s ratios changes in a different way. By applying the [110]-direction strain from 0 to 10$\%$, the relative strain $\varepsilon[1\bar{1}0]$ first decreases and reaches the minimum when $\varepsilon[110]=2\%$. Intriguingly, then $\varepsilon[1\bar{1}0]$ increases until $\varepsilon[110]=10\%$, showing unusual negative Poisson’s ratios.

The strain-induced NPR of this type of material can be explained by the competition between the Au-Au and Au-Te atomic interactions \cite{Pan2020npr}. We can analyze the mechanism with the help of Fig. \ref{figure7}. Without considering the Au-Te atomic interaction, under the external tensile strain F$_{d}$ in the diagonal direction, the Au-Au bond length increases, and the Au$_{3}$ and Au$_{4}$  will move in the D$_{p}$ direction, resulting in a positive Poisson's ratio (PPR). Considering the Au-Te atomic interaction and without considering the interaction between Au-Au atoms, with F$_{d}$, the Te atom will move downwards, and the repulsive interaction will force the atom Au$_{3}$ and Au$_{4}$ move in the D$_{n}$ direction, resulting in NPR property. Combining the two simple phenomena, for the s(I) structure with $\epsilon_{[110]}<2\%$, the attractive interaction among Au atoms play a dominant role, resulting in PPR. For the s(I) structure with $\epsilon_{[110]}>2\%$, the repulsion interaction between Au and Te atoms plays a dominant role, resulting in NPR.

\begin{figure}[htbp]
\includegraphics[scale=1.0]{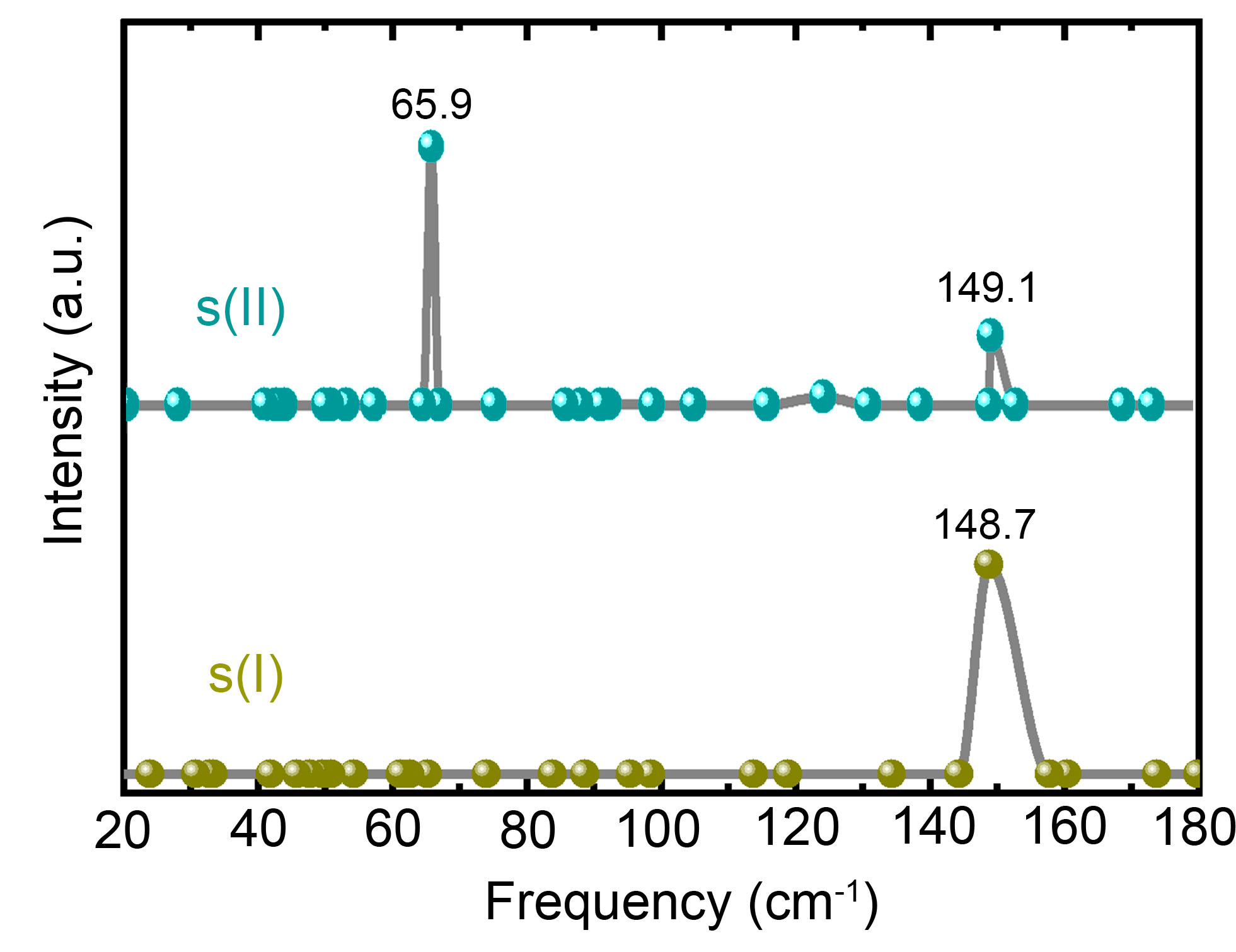}
\caption{The theoretically calculated Raman spectra of s(I)- and s(II)-Au$_{2}$Te. The exact frequencies corresponding to the peaks are shown.}
\label{figure8}
\end{figure}

\subsection{Characterization with Raman spectra}
Characterization of the structures is an important part of the studies of PTMs. A good characterization method can distinguish similar phases and provide evidence of the phase transition in the experiment. Raman spectroscopy is widely used in this field \cite{doi:10.1021/acs.jpclett.0c00256}. In the study of MoTe$_{2}$ by Wang \& Xiao et al. \cite{Wang2017}, the Raman spectra have clearly shown the difference between the 2H phase and the 1T$^{\prime}$ phase of MoTe$_{2}$, and have further shown the phase transition process. In the work of Wang \& Liu et al. \cite{doi:10.1002/adma.201800074}, Raman spectra have shown the phase transition process between the charge density wave (CDW) phase and 1T phase TaS$_{2}$.

In this work, based on first-principles calculations combining with vasp\_raman.py code \cite{vasp_raman_py,doi:10.1063/1.4936965,Wang2017}, we have also computed and compared the Raman spectra of the s(I) and the s(II) phases of Au$_{2}$Te, as shown in Fig. \ref{figure8}. In the Raman spectrum of s(I)-Au$_{2}$Te, there is one prominent peak at 148.7 cm$^{-1}$ whereas for s(II)-Au$_{2}$Te, there are two significant peaks: one is at 65.9 cm$^{-1}$, and the other one is at 149.1 cm$^{-1}$. The most apparent difference between these two spectra is the appearence of a peak at about 65.9 cm$^{-1}$ for s(I) phase, corresponding phonon mode is visualized and shown in the SM (the movie file Raman.mp4). This mode is actually corresponding to the phase transition from s(II) to s(I), which can be softened by diagonal strain, as shown in Fig. \ref{figure3}(b). Thus the Raman spectroscopy is proved to be an excellent method to distinguish the two phases of Au$_{2}$Te.

\section{Summary}
In summary, we systematically investigated the 2D forms of Au$_{2}$Te and predicted two stable Au$_{2}$Te phases, s(I)- and s(II)-Au$_{2}$Te, which are similar in both geometries and potential energies. Their energetic stability is studied and the mechanical, dynamical, and thermal stability is confirmed. Notably, the two phases have quite different electronic and mechanical properties. S(I)-Au$_{2}$Te monolayer is a direct-gap semiconductor, with a gap of 1.252 eV. The carrier mobilities of s(I)-Au$_{2}$Te monolayer are as high as $3.45 \times 10^{4} cm^{2}/(Vs)$ (electron) and $6.47 \times 10^{3} cm^{2}/(Vs)$ (hole), superior to black phosphorus and 1H-MoS$_{2}$. In contrast, s(II)-Au$_{2}$Te monolayer is a topological insulator with ${\mathbb{Z}}_{2}$=1. The nontrivial in-direct bandgap is 28.4 meV, which makes s(II)-Au$_{2}$Te a room-temperature topological insulator. By applying strain along the [110] direction, unusual in-plane negative Poisson ratio can be achieved in s(I)-Au$_{2}$Te, but in s(II)-Au$_{2}$Te, the Poisson ratio is always positive. Most interestingly, the structural phase transition between the semiconducting phase and the topological insulator phase can be achieved with the help of commonly used methods in experiments. By applying tensile strain, a s(II)$ \rightarrow $s(I) phase transition can be achieved. On the other hand, since s(II)-Au$_{2}$Te has lower energy than s(I), a s(I)$ \rightarrow $s(II) phase transition can be expected by chemical means. Moreover, it is found that the electric field can significantly change the energetic relationship of s(I)- and s(II)-Au$_{2}$Te,  which indicates that the s(I)$ \rightarrow $s(II) phase transition can also be achieved by electrostatic gating with the assistance of heat. The results above show that s(I/II)-Au$_{2}$Te monolayers are a new class of 2D materials exhibiting transformation between topological insulator and semiconducting phases, which could be further explored for using in phase-change electronic devices.

\begin{acknowledgments}
We thank Yuhang Liu, Duo Wang, and Chin Shen Ong for their help and useful discussions. This work is supported by the project grant (2016-05366) and Swedish Research Links programme grant (2017-05447) from Swedish Research Council. Linyang Li acknowledges financial support from the National Natural Science Foundation of China (Grant No. 12004097), the Natural Science Foundation of Hebei Province (Grant No. A2020202031), and the Foundation for Introduction of Overseas Scholars of Hebei Province (Grant No. C20200313). Xin Chen thanks China scholarship council for financial support (No. 201606220031). We also acknowledge SNIC-UPPMAX, SNIC-HPC2N and SNIC-NSC centers under the Swedish National Infrastructure for Computing (SNIC) resources for the allocation of time in high-performance supercomputers. Moreover, supercomputing resources from PRACE DECI-15 project DYNAMAT are gratefully acknowledged.
\end{acknowledgments}

\bibliography{Ref}% Produces the bibliography via BibTeX.

\end{document}